\documentclass[a4paper,11pt]{article}

\usepackage{amsmath,amsfonts,amsthm,amssymb}

\newtheorem{Lemma}{Lemma}
\newtheorem{theorem}{Theorem}
\newtheorem{definition}{Definition}
\newtheorem{conjecture}{Conjecture}
\newtheorem{proposition}{Proposition}
\newcommand{\comment}[1]{}

\def\proof{\noindent{\bf Proof:~}}
\def\qed{}

\begin{document}

\title{New separation between $s(f)$ and $bs(f)$}

\author{
Andris Ambainis\thanks{Faculty of Computing, University of Latvia,
Raina bulv. 19, Riga, LV-1586, Latvia, {\tt ambainis@lu.lv}.
Supported by ESF project 1DP/1.1.1.2.0/09/APIA/VIAA/044,
FP7 Marie Curie Grant PIRG02-GA-2007-224886 and
FP7 FET-Open project QCS.} \and Xiaoming Sun\thanks{Institute for Advanced Study, Tsinghua University, Beijing, 100084, China. xiaomings@tsinghua.edu.cn.}}

\date{}
\maketitle

\begin{abstract}
In this note we give a new separation between sensitivity and block sensitivity of Boolean functions: $bs(f)=\frac{2}{3}s(f)^2-\frac{1}{3}s(f)$.
\end{abstract}

\section{Introduction}

Sensitivity and block sensitivity are two commonly used complexity
measures for Boolean functions.
Both complexity measures were originally introduced for
studying the time complexity of CRAW-PRAM's~\cite{CD82,CDR86,Nisan91}.
Block sensitivity is polynomially related to a number of other complexity measures,
including the decision-tree complexity,
the certificate complexity, the polynomial degree, and the quantum query complexity, etc.
(An excellent survey on these complexity
measures and relations between them is~\cite{BW02}.)

A longstanding open problem is the relation between the two measures.
From the definitions of sensitivity and block sensitivity,
it immediately follows that $s(f)\leq bs(f)$ where $s(f)$ and $bs(f)$ denote
the sensitivity and the block sensitivity of a Boolean function $f$.
Nisan and Szegedy~\cite{NS92} conjectured that the sensitivity complexity is also polynomially related to the block sensitivity complexity:

\begin{conjecture}
For every Boolean function f, $bs(f)\leq s(f)^{O(1)}$.
\end{conjecture}

This conjecture is still widely open and the best separation so far is quadratic.
Rubinstein \cite{Rubinstein95} constructed a Boolean function $f$ with
$bs(f)=\frac{1}{2}s(f)^2$ and Virza \cite{Virza10} improved this to
$bs(f)=\frac{1}{2}s(f)^2+\frac{1}{2}s(f)$.

In this paper, we improve this result by constructing a function $f$ with $bs(f)=\frac{2}{3}s(f)^2-\frac{1}{3}s(f)$.

More background and discussion about Conjecture 1 can be found on ~Aaronson's blog~\cite{Aaronson10} and Hatami et al.~\cite{HKP11} survey paper.

\section{Technical preliminaries}

Sensitivity complexity was first introduced by Cook, Dwork and Reischuk~\cite{CD82,CDR86} (under the name critical complexity) for studying the time complexity of CRAW-PRAM's. Let $f:\{0,1\}^n\rightarrow \{0,1\}$ be a Boolean function.
For an input $x\in \{0,1\}^n$, $x^{(i)}$ denotes the input obtained
by flipping the $i$-th bit of $x$. $f^{-1}(1)=\{x|f(x)=1\}$, $f^{-1}(0)=\{x|f(x)=0\}$.

\begin{definition}
\cite{CD82,CDR86}
The {\it sensitivity} complexity of $f$ on input $x$ is defined as $s(f,x)=|\{i|f(x)\neq f(x^{(i)})\}|$. The $0$-{\em sensitivity} and $1$-{\em sensitivity} of the function $f$ is defined as
$$s_0(f)=\max_{x\in f^{-1}(0)}s(f,x), \ \ s_1(f)=\max_{x\in f^{-1}(1)}s(f,x).$$
The {\em sensitivity} is defined as $s(f)=\max\{s_0(f),s_1(f)\}$.
\end{definition}

Nisan~\cite{Nisan91} introduced the concept of block
sensitivity and proved tight bounds for computing $f$ on a CREW-PRAM in terms of block sensitivity.

\begin{definition}
\cite{Nisan91}
The \textit{block sensitivity} of $f$ on input $x$ is the maximum number
$b$ such that there are pairwise disjoint subsets $B_1,\ldots,B_b$
of $[n]$ for which $f(x)\neq f(x^{(B_i)})$, here $x^{(B_i)}$ is the
input obtained by flipping all the bits $x_j$ that $j\in B_i$. We
call each $B_i$ a block. The $0$-{\em block sensitivity} and $1$-{\em block sensitivity} of the function $f$ is defined as
$$bs_0(f)=\max_{x\in f^{-1}(0)}bs(f,x), \ \ bs_1(f)=\max_{x\in f^{-1}(1)}bs(f,x).$$
 The {\em block sensitivity} is defined as  $bs(f)=\max\{bs_0(f),bs_1(f)\}$.
\end{definition}

\section{Previous constructions}

\noindent {\bf Rubinstein's construction}
In~\cite{Rubinstein95} Rubinstein constructed the following composed function $f:\{0,1\}^{4m^2}\rightarrow \{0,1\}$:
$$f(x_{11},\ldots,x_{2m,2m})=\bigvee_{i=1}^{2m}g(x_{i,1},\ldots,x_{i,2m}),$$
  where the function $g:\{0,1\}^{2m}\rightarrow \{0,1\}$ is defined as follows:
\begin{equation*}
g(y_1,\ldots,y_{2m})=1\Leftrightarrow \exists j\in [m], y_{2j-1}=y_{2j}=1, \mbox{and } y_k=0~(\forall k\notin \{2j-1,2j\})
\end{equation*}

It is not hard to see that for the function $f$, $s(f)=2m$ and $bs(f)=2m^2$, so $bs(f)=\frac{1}{2}s(f)^2$.

\vskip10pt

\noindent {\bf Virza's construction}
Recently Virza~\cite{Virza10} slightly improved this separation by constructing a new function $f:\{0,1\}^{(2m+1)^2}\rightarrow \{0,1\}$:
$$f(x_{11},\ldots,x_{2m+1,2m+1})=\bigvee_{i=1}^{2m+1}g(x_{i,1},\ldots,x_{i,2m+1}),$$
 where the function $g:\{0,1\}^{2m+1}\rightarrow \{0,1\}$ is defined as follows:
\begin{eqnarray*}
g(y_1,\ldots,y_{2m+1})=1 &\Leftrightarrow &
\left( \exists j\in[m]~y_{2j-1}=y_{2j}=1~\mbox{and}~\forall k\notin\{2j-1,2j\}~y_k=0\right)\\
& &\mbox{ or }
\left(y_{2m+1}=1~\mbox{and}~\forall j\neq 2m+1~y_j=0\right)
\end{eqnarray*}

It can be verified that 
$s(f)=2m+1$ and $bs(f)=(2m+1)(m+1)$, so $bs(f)=\frac{1}{2}s(f)^2+\frac{1}{2}s(f)$.

\vskip15pt
Rubinstein's and Virza's constructions both use the same strategy, constructing the
function $f$ by composing OR (on the top level) with a function $g$ (on the bottom level).
In this paper, we systematically explore the power of this strategy.

In the next section, we characterize the sensitivity and the block sensitivity
of functions obtained by such composition.
In Section~\ref{section:2/3}, we improve the constant $c$ in the separation
$s(f)=c\cdot bs^2(f)$ from $\frac{1}{2}$ to $\frac{2}{3}$.
In Section~\ref{section:optimal}, we show that $s(f)=(\frac{2}{3}+o(1)) bs^2(f)$
is optimal for functions obtained by composing OR with a function $g$ for which $s_0(g)=1$.

\section{Separations between $s(f)$ and $bs(f)$ for composed functions}

We consider functions $f$ obtained by composing OR with a function $g$.
\begin{equation}
\label{eq:or}
f(x_{11},\ldots,x_{n,m})=\bigvee_{i=1}^{n}g(x_{i,1},\ldots,x_{i,m}),
\end{equation}
We have

\begin{Lemma}
\label{lem:compose}
\begin{enumerate}
\item[(a)]
$s_0(f) = n \cdot s_0(g)$;
\item[(b)]
$s_1(f) = s_1(g)$.
\item[(c)]
$bs_0(f) = n \cdot bs_0(g)$;
\end{enumerate}
\end{Lemma}

\proof
Part (a):
Let $x=(x_1, \ldots, x_m)$ be the input on which $g(x_1, \ldots, x_m)$ achieves
the maximum 0-sensitivity $s_0(g)$. Then, $g(x)=0$ but there exist $s_0(g)$ distinct
$j_1, \ldots, j_{s_0(g)}\in [m]$ for which $g(x^{(j_l)})=1$ ($l\in[s_0(g)]$).

We consider the input $y=(y_{11}, \ldots, y_{nm})$ for the function $f$
obtained by replicating $x$ $n$ times: $y_{1j}=y_{2j}=\ldots=y_{nj}=x_j$.
Then, $f(y)=0$ but $f(y^{(i, j_l)})=1$ for any $i\in[n]$, $l\in[s_0(g)]$.
Thus, $s_0(f)\geq n \cdot s_0(g)$.

Conversely, assume that $f(y_{11}, \ldots, y_{nm})$
achieves sensitivity $s_0(f)$ on an input $y=(y_{11}, \ldots, y_{nm})$.
Then, there exists $i\in[n]$ such that there are at least
$\frac{s_0(f)}{n}$ sensitive variables among $y_{i1}, \ldots, y_{im}$.
We take the input $x=(x_1, \ldots, x_n)$ for $g$ defined by
$x_j = y_{ij}$. Then, $g(x)=0$ 
and $g(x^{(j)})=1$ for all variables $j$ such that $y_{ij}$ is sensitive for
$g$ on the input $y$.
Hence, $s_0(g)\geq \frac{s_0(f)}{n}$.

Part (b): For $s_1(f)\geq s_1(g)$, let $x=(x_1, \ldots, x_m)$ be the
input on which $g$ achieves the maximal 1-sensitivity and let
$x'=(x'_1, \ldots, x'_m)$ be any input with $g(x')=0$. We define
$y=(y_{11}, \ldots, y_{nm})$ by $y_{1i}=x_i$ and $y_{2i}=\ldots=y_{ni}=x'_i$ ($i\in [m]$).
Then, $f(y)=g(x)=1$ and $f(y^{(1j)})=g(x^{(j)})=0$ for all variables $j$ such that $x_{j}$ is sensitive for
$g$ on the input $x$.
Hence, the sensitivity of $f$ on $y$ is at least the sensitivity of $g$ on $x$.

For $s_1(f)\leq s_1(g)$, we assume that $f(y)$
achieves the maximum sensitivity $s_1(f)$ on an input $y=(y_{11}, \ldots, y_{nm})$.
Then, it must be the case that $g(y_{i1}, \ldots, y_{im})=1$ for exactly
one $i$.
\comment{(If $g(y_{i1}, \ldots, y_{im})=0$ for all $i$, then, by (\ref{eq:or}),
$f(y_{11}, \ldots, y_{nm})=0$. If $g(y_{i1}, \ldots, y_{im})=1$ for more than
one $i$, then, after changing any one variable $y_{ij}$, there will be at
least one $i'$ remaining with $g((y_{i'1}, \ldots, y_{i'm})=1$.
Hence, $f$ would not be sensitive to any variable $y_{ij}$ on the input $y$.)}
Moreover, if $i'\neq i$, then $f(y^{(i',j)})=1$ and $f$ is not sensitive to
changing $y_{i'j}$.

Let $x_1=y_{i1}$, $\ldots$, $x_m=y_{im}$. Then, $f(y^{(ij)})=0$ if and only if
$g(x^{(j)})=0$. Hence, the sensitivity of $f$ on the input $y$ is equal to the sensitivity
of $g$ on the input $x$. This means that $s_1(g)\geq s_1(f)$.

The proof of part (c) is similar to the proof of part (a).
\qed

\section{a $\frac{2}{3}$-separation}\label{section:2/3}

\begin{theorem}\label{theorem1}
For any $m\in \mathbb{N}$, there is a Boolean function $f$ on $(4k+2)(3k+2)$ variables, such that $s(f)=3k+2$, $bs(f)=(3k+2)(2k+1)$, thus
$bs(f)=\frac{2}{3}s(f)^2-\frac{1}{3}s(f)$.
\end{theorem}

\proof  Suppose $n=2(2k+1)$ here.
Define $g:\{0,1\}^n\rightarrow \{0,1\}$ as follows:
\begin{eqnarray*}
g(y_1,\ldots,y_n)=1 &~\Leftrightarrow~& \exists j\in[2k+1]~ \left( x\mbox{ satisfies pattern } P_j\right),
\end{eqnarray*}
where pattern $P_j$ ($j=1,\ldots,2k+1$) is defined as
\begin{equation*}
P_j: ~ x_{2j-1}=x_{2j}=1,\mbox{ and } \forall i\in [m], x_{2j+2i}=0,x_{2j-2i-1}=x_{2j-2i}=0.
\end{equation*}
Here the index of $x_*$ is modular $n$. We use the notation $x\sim P$ to represent $x$ satisfies pattern $P$.

\begin{proposition}\label{proposition1}
$s_1(g)=3k+2$, $s_0(g)=1$, and $bs_0(g)=n/2=2k+1$.
\end{proposition}

\noindent {\it Proof of Proposition~\ref{proposition1}.} For any $x\in g^{-1}(1)$, by definition there exists $j\in [2k+1]$, such that $x\sim P_j$.
The bits in pattern $P_j$ form a {\em certificate} of $x$, and it contains all the possible sensitive bits of $x$. Thus $s(f,x)\leq 2+3k$. On the other hand
$f(110\ldots 0)=1$, and $s(f,110\ldots 0)=3k+2$. Therefore, $s_1(f)=3k+2$.

Since $f(0\ldots 0)=0$, $f(110\ldots0)=f(00110\ldots0)=\cdots=f(0\ldots 011)=1$, so $bs(f,0\ldots 0)\geq n/2=2k+1$, thus $bs_0(f)\geq 2k+1$. This is already enough for our purpose, but for completeness we will show $bs_0(f)\leq 2k+1$. For any $x\in g^{-1}(0)$, suppose $bs(g,x)=b$ and $B_1,\ldots,B_b$ be minimal pairwise disjoint blocks so that $g(x^{(B_i)})=1$ ($i\in [b]$). By the definition of $g$, for each $B_i$ there exists a $j\in [2k+1]$,
$x^{(B_i)}\sim P_j$. Since $B_1,\ldots,B_b$ are pairwise disjoint, it is easy to see that different $B_i$ corresponds to different $P_j$. Therefore, $b\leq 2k+1$.

Next we show that $s_0(g)=1$. Suppose there exists $x\in \{0,1\}^n$, $g(x)=0$ and $s_0(g,x)\geq 2$, i.e. $\exists~i\neq i'\in [n]$, $g(x^{(i)})=g(x^{(i')})=1$, by the definition of $g$, there are $j,j'\in [2k+1]$, $x^{(i)}\sim P_j$ and $x^{(i')}\sim P_{j'}$. Since $i\neq i'$ and $g(x)=0$, it is easy to see that $j\neq j'$. We claim that for any $y\sim P_j$ and any $z\sim P_{j'}$, the Hamming distance between $y$ and $z$ $h(y,z)\geq 3$. But it is clear that $h(x^{(i)},x^{(i')})\leq 2$, contradiction.

 W.l.o.g. we assume $j<j'$, consider the value of $j'-j$, there are two cases:

\begin{enumerate}
  \item If $j'-j\leq k$: let's consider the three coordinates $2j-1,2j$ and $2j'$, since $y\sim P_j$, by definition $y_{2j-1}=1$, $y_{2j}=1$, and $y_{2j'}=y_{2j+2(j'-j)}=0$. On the other hand $z\sim P_{j'}$, so $z_{2j'}=1$, $z_{2j-1}=z_{2j'-2(j'-i)-1}=0$, and $z_{2j}=z_{2j'-2(j'-j)}=0$. Hence $h(y,z)\geq 3$.

  \item If $j'-j>k$: we consider the three coordinates $2j,2j'-1$ and $2j'$ in this case. Since $y\sim P_j$, so $y_{2j}=1$, $y_{2j'-1}=y_{2j-2(n/2+j-j')-1}=0$, and $y_{2j'}=y_{2j-2(n/2+j-j')}=0$, here we use the property that the index is modular $n$.
      $z\sim P_{j'}$ implies that $z_{2j'-1}=z_{2j'}=1$, and $z_{2j}=z_{2j'-2(j'-j)}=0$. Therefore, $h(y,z)\geq 3$.
\end{enumerate}

This complete the proof of Proposition~\ref{proposition1}. \hfill$\Box$

\comment{
\vskip15pt

\noindent {\it Proof of Theorem~\ref{theorem1} (continue)}
Let's consider the following composition function $f:\{0,1\}^{(3m+2)(4m+2)}\rightarrow \{0,1\}$:
$$f(x_{1,1},\ldots,x_{1,n},x_{2,1},\ldots,x_{(3m+2),n})=\bigvee_{i=1}^{3m+2}g(x_{i,1},\ldots,x_{i,n}).$$
then
$$s_0(f)=(3m+2)s_0(g)=3m+2, \ s_1(f)=s_1(g)=3m+2.$$ Thus $s(f)=3m+2$.
$$bs_0(f)=(3m+2)bs_0(g)=(3m+2)(2m+1),\ bs_1(f)=bs_1(g)=3m+2.$$
Thus $bs(f)=(3m+2)(2m+1)$,
\begin{equation*}
bs(f)=\frac{2}{3}s(f)^2-\frac{1}{3}s(f).
\end{equation*}
\hfill
}

Theorem \ref{theorem1} follows by applying Lemma \ref{lem:compose} with
$n=3m+2$ to the function $g$ of Proposition \ref{proposition1}.
$\Box$

\section{The optimality of 2/3 example}\label{section:optimal}

We claim that the 2/3 example is essentially
optimal, as long as we consider functions $g$ with $s_0(g)=1$.

\begin{theorem}
\label{theorem2}
Assume that we have a function $g$ with $s_0(g)=1$ and $bs_0(g)=k$.
Then, $s_1(g)\geq 3 \frac{k-1}{2}$.
\end{theorem}

Given such function $g$, we can obtain the biggest separation when we use
Lemma \ref{lem:compose} with $n=s_1(g)$. Then, $s_0(f)=n=s_1(g)$, $s_1(f)=s_1(g)$ and
\[ bs_0(f) = n \cdot bs_0(g) = s_1(g) \cdot k \leq
s_1(g) \left( \frac{2}{3} s_1(g) + 1 \right) .\]

\proof
Without the loss of generality, we can assume that the maximum sensitivity is achieved
on the all-0 input which we denote by $0$. Let $B_1, \ldots, B_k$ be the sensitive blocks.
We assume that each $B_i$ is minimal (i.e., $f$ is not sensitive to changing variables in
any $B'\subset B_i$).

Since $s_0(g)=1$, $g$ must have the following structure: $g(x)=1$
iff $x$ belongs to one of several subcubes $S_i$ defined in a following way:
\begin{equation}
\label{eq:si}
S_i = \{ (x_1, \ldots, x_N) | x_{i_1}=\ldots =x_{i_l}=0, x_{j_1}=\ldots =x_{j_m}=1 \} ,
\end{equation}
with any two inputs $(x_1, \ldots, x_N)$, $(y_1, \ldots, y_N)$ belonging to
different $S_i$'s differing in at least 3 variables.

The inputs $0^{(B_i)}$ all belong to different $S_i$'s, since $0^{(B_i)}\in S_l$,
$0^{(B_j)}\in S_l$ would imply $0\in S_l$ and $g(0)=1$.
We assume that $0^{(B_1)}\in S_1$, $\ldots$, $0^{(B_k)}\in S_k$.
We can assume that there is no other subcubes $S_i$.
(Otherwise, we can replace $g$ by $g'$, $g'(x)=1$ if $x\in \cup_{i=1}^k S_i$.)
For a subcube (\ref{eq:si}), we denote $I_i=\{i_1, \ldots, i_l\}$,
$J_i = \{ j_1, \ldots, j_m\}$.

Since $0^{(B_i)} \in S_i$, we must have $J_i\subseteq B_i$.
Moreover, we also have $g(0^{(B')})=0$ for any $B'\subset B_i$.
Hence $0^{(B')}\notin S_i$ for any such $B'$.
This means that $J_i=B_i$.

If $s_0(g)=1$, then any $x\in S_i$ and $y\in S_j$, $i\neq j$ must differ in
at least 3 variables. This means that
\[ \left| (I_i \cap J_j) \cup (J_i \cap I_j) \right| \geq 3 .\]
Hence,
\[ \sum_{i, j: i\neq j} | I_i \cap J_j | \geq 3 \frac{k(k-1)}{2} .\]
This means that, for some $i$,
\[ \sum_{j} | I_i \cap J_j | \geq 3 \frac{k-1}{2} .\]
Since $J_j=B_j$ and blocks $B_j$ are disjoint,
this means that $|I_i|\geq 3 \frac{k-1}{2}$.

For an input $x\in S_i$, changing any variable in $I_i$ results in an input $y\notin S_i$.
Hence, $x\in S_i$ is sensitive to all $j\in I_i$
and $s_1(g)\geq 3 \frac{k-1}{2}$.
\qed

\section{Conclusion and Discussion}

We have improved the best separation between the sensitivity and the block sensitivity
from $bs(f)=\frac{1}{2}s(f)^2+\frac{1}{2}s(f)$
to $bs(f)=\frac{2}{3}s(f)^2-\frac{1}{3}s(f)$.

The obvious open question is whether further improvements are possible,
using the same strategy of composing OR with a cleverly chosen function $g$.
If such improvements are possible, they must use functions
$g$ with $s_0(g)>1$ (because of Theorem \ref{theorem2}).

\end{document}